\documentclass[a4paper,fleqn]{cas-sc}

\usepackage[numbers,sort&compress]{natbib}

\usepackage{tabularx}

\newcommand{\rev}[1]{#1}
\newcommand{\nrev}[1]{#1}

\newcolumntype{L}[1]{>{\raggedright\arraybackslash}p{#1}}
\newcolumntype{C}[1]{>{\centering\arraybackslash}p{#1}}

\begin{document}
\let\WriteBookmarks\relax
\def\floatpagepagefraction{1}
\def\textpagefraction{.001}

\shorttitle{Verification-driven closed-loop multi-agent structural design}
\shortauthors{J. Luo et~al.}

\title[mode=title]{Verification-driven closed-loop multi-agent large language model framework for code-compliant structural design}

\author[1]{Jianbin Luo}[orcid=0000-0002-2244-0440]
\cormark[1]
\ead{ljb312@fzu.edu.cn}
\credit{Conceptualization, Methodology, Validation, Writing -- original draft, Project administration, Funding acquisition}

\affiliation[1]{organization={College of Civil Engineering, Fuzhou University},
            city={Fuzhou},
            postcode={350108},
            country={China}}

\author[1]{Weibin Lin}
\credit{Writing -- original draft}

\author[1]{Yiran Lin}
\credit{Software, Data curation, Validation}

\author[1]{Qing Wei}
\credit{Investigation, Resources, Software}

\author[1]{Wei Guo}
\credit{Writing -- original draft, Writing -- review \& editing}

\cortext[1]{Corresponding author}

\begin{abstract}
\rev{Multi-agent large language model (LLM) systems are applied to structural design, yet most use one-shot generation and cannot verify their output, leaving them ill-suited to safety-critical tasks.} \rev{Rather than trusting LLM self-correction, this framework injects feedback from an external physics-based verifier into a closed repair loop.} \rev{The framework couples a three-layer finite-element verification system with a dual-node loop\nrev{. Node~1} turns code violations into hard repair constraints, Node~2 turns a four-dimensional quality score into safety-first soft constraints, and a retrieval-augmented code base makes every violation traceable to a clause.} Over five structure types and 44 cases, code compliance rises from 56.8\% to 98.6\% and the composite score from 63.8 to 71.4 ($p<10^{-6}$), using about 5.8\% less material. Removing either node degrades performance, and \rev{compliance does not change detectably across the two backbone LLMs tested, indicating that it is here attributed to the external verifier rather than the model.} \rev{The framework, the 44-case benchmark and all experiment scripts are released as open source for replicability.}
\end{abstract}


\maketitle

\section{Introduction}

\subsection{Background}
Structural design is a process of iteratively trading off multiple objectives, such as safety and economy, under code constraints. CAD, BIM and general-purpose finite-element software (SAP2000, ANSYS, OpenSees, etc.) are already mature for information modelling and structural analysis \cite{sun2021}, but the steps that depend most heavily on professional judgement \nrev{(translating design intent into parameters, interpreting code clauses and reaching a verdict, and reconciling computational results across different software)} still have to be carried out manually. \nrev{These steps are the bottleneck for automating structural design. Iteration speed is limited by the engineer}, output quality varies with experience, and repeated cross-software data re-entry readily introduces new errors.

Large language models offer a way past this bottleneck. A large language model (LLM) can interpret natural language, draw on domain knowledge and orchestrate tools, and could in principle act as a coordinator for intent interpretation, tool invocation and result integration \cite{chiarello2024}. Applying an LLM directly to structural design, however, runs into a fundamental difficulty\nrev{. It} is a language model rather than a mechanics solver \cite{gopfert2024}, so the schemes it produces are often coherent and plausible in wording yet may violate mechanical principles or code limits\nrev{, the familiar problem of hallucination.} In safety-critical design, a result that has not been independently verified is not fit for engineering use. More importantly, recent studies of LLM self-correction reach a strikingly consistent negative conclusion\nrev{. Without} external feedback, self-correction on reasoning tasks is unreliable and can even turn a correct answer into a wrong one \cite{kamoi2024,huang2024}. Prompt engineering or model self-reflection alone, therefore, can hardly guarantee the reliability of engineering-design results, \nrev{and} effective automatic repair requires an external verifier that supplies an objective signal \cite{madaan2023}. Building on this premise, the present work uses a finite-element solver and a code checker as an arbitration mechanism independent of the LLM, which provides the feedback signal that drives the closed loop. Our central hypothesis is that code compliance in LLM-driven structural design need not come from a more capable model, but from closing the generation--verification loop with an external verifier\nrev{. The} experiments bear this out, with an open-loop baseline plateauing at a 56.8\% compliance rate while the same agents inside a verification-driven loop reach 98.6\%.

\subsection{State of the art and research gaps}\label{sec:gaps}
Research on intelligent structural design has moved from searching for optimal solutions \cite{rajeev1992,kicinger2005} to learning to generate them with GANs, graph neural networks, variational autoencoders and deep reinforcement learning \cite{liao2021,lu2022,zhao2023a,zhao2023b,mirra2021,gao2024}, and, over the past two years, to multi-agent LLM frameworks such as the code-compliant designer of Chen and Bao \cite{chen2025} and the tool-coupled MASSE system \cite{liang2025}. In parallel, LLM-driven code-compliance checking and retrieval-augmented generation (RAG) have anchored LLM output to code text \cite{eastman2009,yang2024,shi2025,ying2024,chung2025}. Section~\ref{sec:related-data} reviews this literature in detail.

Nevertheless, from the standpoint of trustworthy engineering AI, the existing literature still shares three gaps:
\begin{enumerate}
\item \textbf{Generation-paradigm gap.} Existing multi-agent frameworks are open-loop: when verification flags a problem, repair falls back on manual intervention or unguided retries, with no automatic loop that structures the verification signal and drives directional repair within an explicit constraint space; and no study decomposes, at the process level, what each closed-loop stage contributes.
\item \textbf{Trustworthiness-verification gap.} Verification has stayed at a single level (code-limit checking or after-the-fact comparison with commercial software), with no scheme covering the full chain of engine accuracy, model construction and result compliance; in particular, the finite-element engine wrapper inside the workflow has never been independently benchmarked, although an engine-layer error would invalidate every higher-level conclusion.
\item \textbf{Interpretability gap.} Violation verdicts are returned as Boolean flags or bare numbers that cannot be traced to a code clause, and the multi-dimensional assessment weights are set opaquely, so neither the AI's decision nor its quality assessment is auditable.
\end{enumerate}

\subsection{Contributions}
Targeting the three gaps above, this work designs and implements a verification-driven closed-loop multi-agent large language model framework for code-compliant structural design. The main contributions are as follows.

\textbf{First, a dual-node closed-loop repair mechanism.} Repair feedback comes solely from an external verifier (finite-element analysis plus code checking), not from LLM self-correction. Node~1 writes violation items and their exceedance magnitudes into the prompt as hard constraints for directional repair; Node~2 maps the four-dimensional score onto safety-first parameter-adjustment constraints and selects among independently verified candidates. Using score-history records we decompose the two nodes' contributions, showing how compliance repair and quality refinement divide the work and where each reaches its limit.

\textbf{Second, a three-layer finite-element trustworthiness-verification system.} It stacks engine-accuracy regression against analytical and independent ANSYS solutions, automatic model-construction checks with geometric-preview confirmation, and structured code checking with safety-factor computation, forming a defence-in-depth against hallucination that can serve as a general verification template for other LLM-driven engineering-computation systems.

\textbf{Third, a traceable RAG code base and an auditable assessment-weight scheme.} Violation diagnosis retrieves and cites GB~50010 \cite{cijk2010} and GB~50017 \cite{gb50017} clauses, and the four-dimensional weights follow a transparent, auditable two-factor scoring procedure.

Beyond these, the framework is delivered as a working system with an extensible factory-style architecture and a natural-language web interface for non-experts (Sections~\ref{sec:framework} and~\ref{sec:impl}); the complete code, test set and scripts are released as open source, as delivery and reproducibility features rather than scientific claims.

\section{Related work}

\subsection{Data-driven intelligent structural design}\label{sec:related-data}
Early research on intelligent structural design centred on casting the design problem as a mathematical optimization model and searching a predefined design space for a constraint-satisfying optimum. Genetic algorithms, simulated annealing and expert systems were applied to structural selection and section optimization \cite{rajeev1992}. Evolutionary methods such as an improved estimation-of-distribution algorithm \cite{kicinger2005} and multi-island genetic algorithms showed strong search ability for discrete-variable problems in truss and industrial-structure optimization. The drawback of these methods is that the design space must be defined by hand and the model rebuilt for every new problem.

Deep learning then shifted the paradigm from searching for optima to learning to generate solutions. StructGAN, proposed by Liao et al. \cite{liao2021}, was the first to apply an image-translation network to shear-wall layout generation, achieving an end-to-end mapping from architectural to structural drawings. Later studies embedded a mechanical-performance evaluator in the loss function \cite{lu2022} or used attention to learn engineering heuristics \cite{zhao2023a}, improving the mechanical rationality and detailing compliance of the generated results. Graph neural networks have been used to learn the topological relations of frame-beam layouts \cite{zhao2023b} and, combined with exploratory genetic algorithms, to optimize steel-reinforcement layouts \cite{li2023}; an AI-generated design space built with variational autoencoders outperformed manually defined variable spaces in both diversity and performance \cite{mirra2021}; physics-informed deep reinforcement learning has produced safe, economical steel-frame designs within seconds \cite{gao2024}; and physics-rule-guided self-supervised GANs have extended generative design to base-isolated shear-wall structures \cite{liao2024}. These methods, however, generally require large amounts of labelled data, generalize only within the training distribution, and offer little interpretability\nrev{. They} can neither justify a particular decision nor guarantee clause-by-clause compliance. Knowledge-driven LLM methods have arisen precisely to address these shortcomings.

\subsection{LLMs and multi-agent systems in engineering design and code compliance}
Analysing 15{,}355 engineering-design papers, Chiarello et al. \cite{chiarello2024} identified three roles an LLM can play in engineering design\nrev{, namely generation, evaluation and description.} G\"opfert et al. \cite{gopfert2024}, however, cautioned that an LLM is fundamentally a language model rather than a calculator and cannot, on its own, take on safety-critical design tasks. Multi-agent frameworks turn the LLM from a conversational tool into a collaborating participant. Chen and Bao \cite{chen2025} proposed a multi-agent code-compliant design framework for reinforced-concrete structures, in which dedicated LLM agents handle query classification, parameter extraction, structural analysis and reinforcement design and produce a verifiable calculation report through inter-agent cross-checking; the MASSE system of Liang et al. \cite{liang2025} reproduces the analyst, engineer and management roles of a consulting team and integrates OpenSees and ANSYS to complete a real rack-system design. The paradigm is spreading quickly to other AEC sub-fields\nrev{. A} parallel line of work develops agentic LLMs for automated structural analysis, progressing from beam analysis \cite{liu2025agent} to 2D frames \cite{geng2025} and across multiple solver platforms \cite{geng2026}, with dedicated architectures introduced to curb hallucination in multi-step modelling \cite{cheng2026}; multi-agent coordination has also been applied to ultra-high-performance concrete design \cite{chenbao2026,guo2025} and, through a router that classifies tasks and selects experts, to foundation design \cite{youwai2026}; and Dong et al. \cite{dong2025} built an LLM-driven multi-agent BIM coordinator for non-expert interaction, while Zhang et al. \cite{zhang2025} introduced reusable LLM-agent patterns and an open-source agent library for building-energy analysis. The same shift runs through the wider engineering-AI literature: large language model multi-agent frameworks have been applied to sustainable industrial design \cite{alevizos2026eaai}, planning agents with generative memory improve long-horizon task performance \cite{liu2025eaai}, and LLM-driven retrieval supports risk identification in underground-space engineering \cite{miao2026eaai}. Together these signal a move toward automation and standardization of LLM agents across engineering.

Research on code intelligence dates back to the automated code compliance checking (ACC) of Eastman et al. \cite{eastman2009}. The arrival of LLMs has sharply lowered the barrier to turning code clauses into computable rules\nrev{. Yang} and Zhang \cite{yang2024} used prompt engineering to automate the transformation of building-code information; the BuildThemis framework \cite{shi2025} pairs a domain-fine-tuned LLM with RAG to generate executable compliance-checking scripts; Ying and Sacks \cite{ying2024} proposed an autonomous compliance-checking framework with the LLM as agent, able to understand design requirements, plan checking tasks and retrieve BIM data; and GraphCompliance \cite{chung2025} represents code text as a policy graph and the runtime context as an event graph of subject--action--object triples, aligns the two, and hands them to the LLM for structured reasoning, improving compliance-judgement accuracy over pure-LLM and plain-RAG baselines in a GDPR setting. These works show that RAG and structured knowledge can anchor LLM output to code text and achieve clause-level traceability, but they remain confined to the checking step and have not been folded into a generation--repair closed loop.

Work on LLM reliability bears directly on the present study. Self-Refine \cite{madaan2023} proposed an iterative generate--feedback--refine scheme, but the survey of Kamoi et al. \cite{kamoi2024} and the experiments of Huang et al. \cite{huang2024} converge on the same finding\nrev{. Without} an external signal, LLM self-correction brings little benefit on reasoning tasks and can even be counterproductive; reliable correction needs feedback from external tools such as a code interpreter, a retrieval system or a dedicated verifier. CRITIC \cite{gou2024} and Self-Debug \cite{chen2024}, which repair code using interpreter-execution results, are cases in point. We carry this idea over to structural design, treating the finite-element solver and the code checker as an arbitration mechanism the LLM cannot influence and using it as the feedback source for the closed loop.

\subsection{Positioning of this work}
Table~\ref{tab:compare} situates this work against the two most closely related studies. The literature has already answered, in the affirmative, whether multi-agent systems can carry out structural design. \nrev{The questions we take up are different. We ask} how a verification signal becomes the driving force for automatic repair, at which step and by how much the closed loop helps, and what underwrites the trustworthiness of the whole process.

\begin{table}[htbp]
\centering
\caption{Comparison of this work with representative studies.}
\label{tab:compare}
\small
\begin{tabularx}{\linewidth}{@{}L{0.16\linewidth} L{0.22\linewidth} L{0.20\linewidth} X@{}}
\toprule
\textbf{Dimension} & \textbf{Chen \& Bao \cite{chen2025}} & \textbf{MASSE \cite{liang2025}} & \textbf{This work} \\
\midrule
Structure types & RC beam/column members & Rack-type steel structures & 5 types (beams/trusses/frames); architecture supports zero-intrusion extension \\
Verification & After-the-fact comparison with SAP2000 & Professional-software integration & Three-layer system (engine benchmark + model check + code check) \\
Feedback mode & Open loop (agent cross-check) & Open loop / manual intervention & Dual-node automatic closed loop \\
Closed-loop mechanism analysis & --- & --- & Two-stage contribution decomposition via score history \\
Code traceability & Code rules embedded in prompt & --- & RAG clause-level automatic citation \\
Assessment dimensions & Compliance & Efficiency / accuracy & Four-dimensional quantification + objective-mechanics cross-validation \\
Weight auditability & --- & --- & Two-factor scoring--normalization method \\
\bottomrule
\end{tabularx}
\end{table}

\section{Methodology}

\subsection{Overall framework}\label{sec:framework}
\begin{figure}[htbp]
\centering
\includegraphics[width=0.82\linewidth]{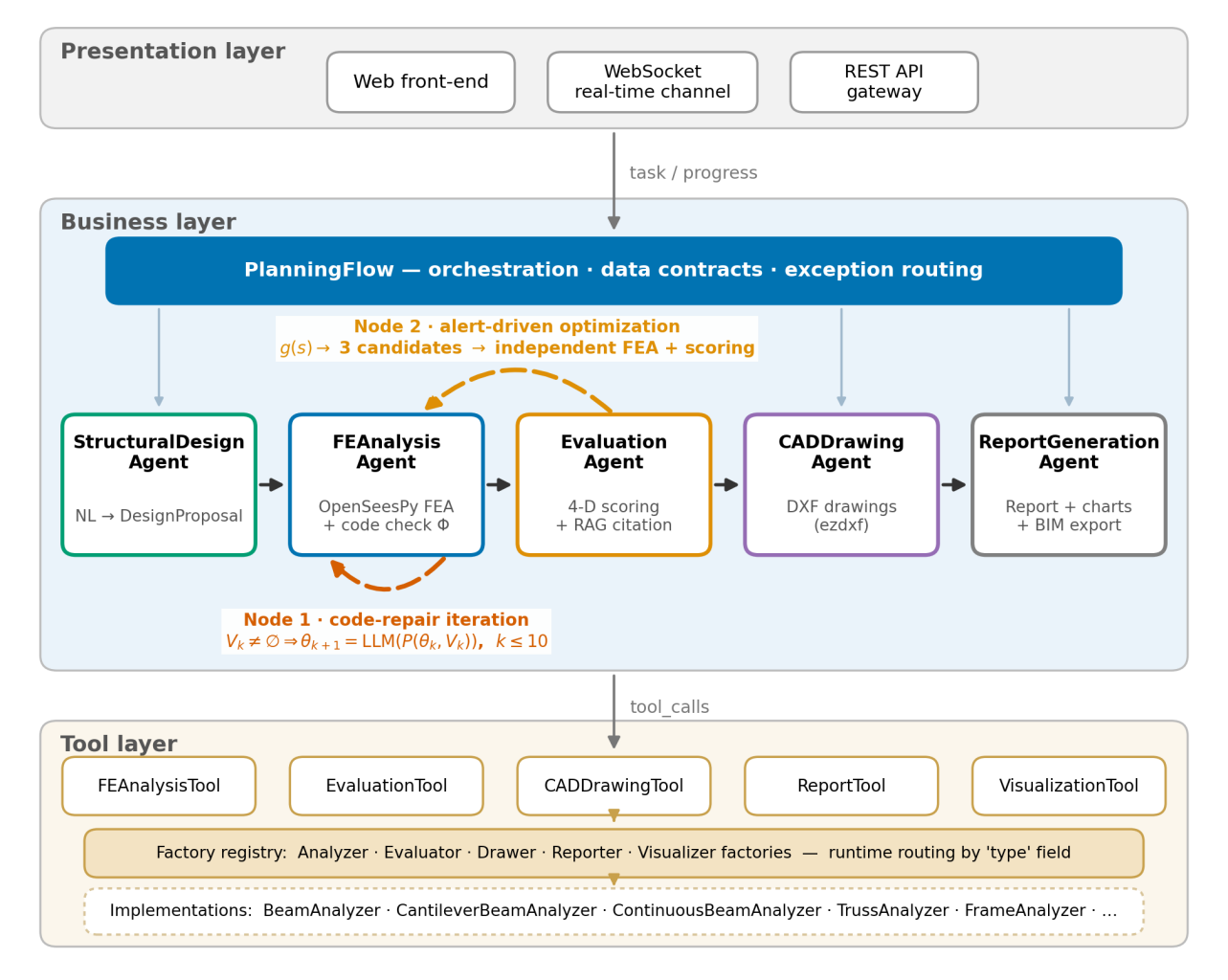}
\caption{Overall architecture of the verification-driven closed-loop multi-agent framework.}
\label{fig:arch}
\end{figure}

The framework casts the structural-design workflow as a five-stage state machine driven by a task orchestrator, on a three-tier architecture of presentation, business and tool layers (Fig.~\ref{fig:arch}). The business layer comprises the orchestrator PlanningFlow and five specialized agents\nrev{, namely} StructuralDesignAgent (requirement understanding and parameterization), FEAnalysisAgent (finite-element analysis and code checking), EvaluationAgent (four-dimensional assessment and RAG code tracing), CADDrawingAgent (drawing generation) and ReportGenerationAgent (report aggregation). Each agent inherits from the ToolCallAgent base class of the OpenManus framework and completes its stage through tool invocation under a ReAct loop \cite{yao2023}; structured data move between stages through explicitly defined JSON contracts (DesignProposal, AnalysisResults, EvaluationReport, DrawingResults, ReportResults), whose field-level schemas are released with the code.

The framework's extensibility rests on the principle of generic agents plus tool routing\nrev{. The} agents hold no structure-type-specific decision logic, and every type-specific implementation (e.g. BeamAnalyzer, TrussDrawer) is pushed down to the tool layer, where five factory classes (AnalyzerFactory and the like) route dynamically at runtime according to the \texttt{type} field of the design scheme. Adding a structure type only requires implementing and registering the corresponding class at the tool layer, leaving the agent code and prompt framework untouched; even the list of supported types in an agent's system prompt is read at runtime from the factory registry. This separation keeps semantic understanding and decision-making in the LLM layer and delegates exact numerical computation to deterministic code, so the risk of hallucination is contained at the boundary between the two.

PlanningFlow acts as both data hub and exception handler between stages\nrev{. Before} the analysis stage it pre-checks with the factory whether the structure type is registered; on analysis failure it re-routes; and when code checking fails or an assessment warning fires, it enters the first or second closed-loop node of Section~\ref{sec:dualnode}, respectively.

\subsection{Three-layer finite-element trustworthiness-verification system}
\begin{figure}[htbp]
\centering
\includegraphics[width=0.95\linewidth]{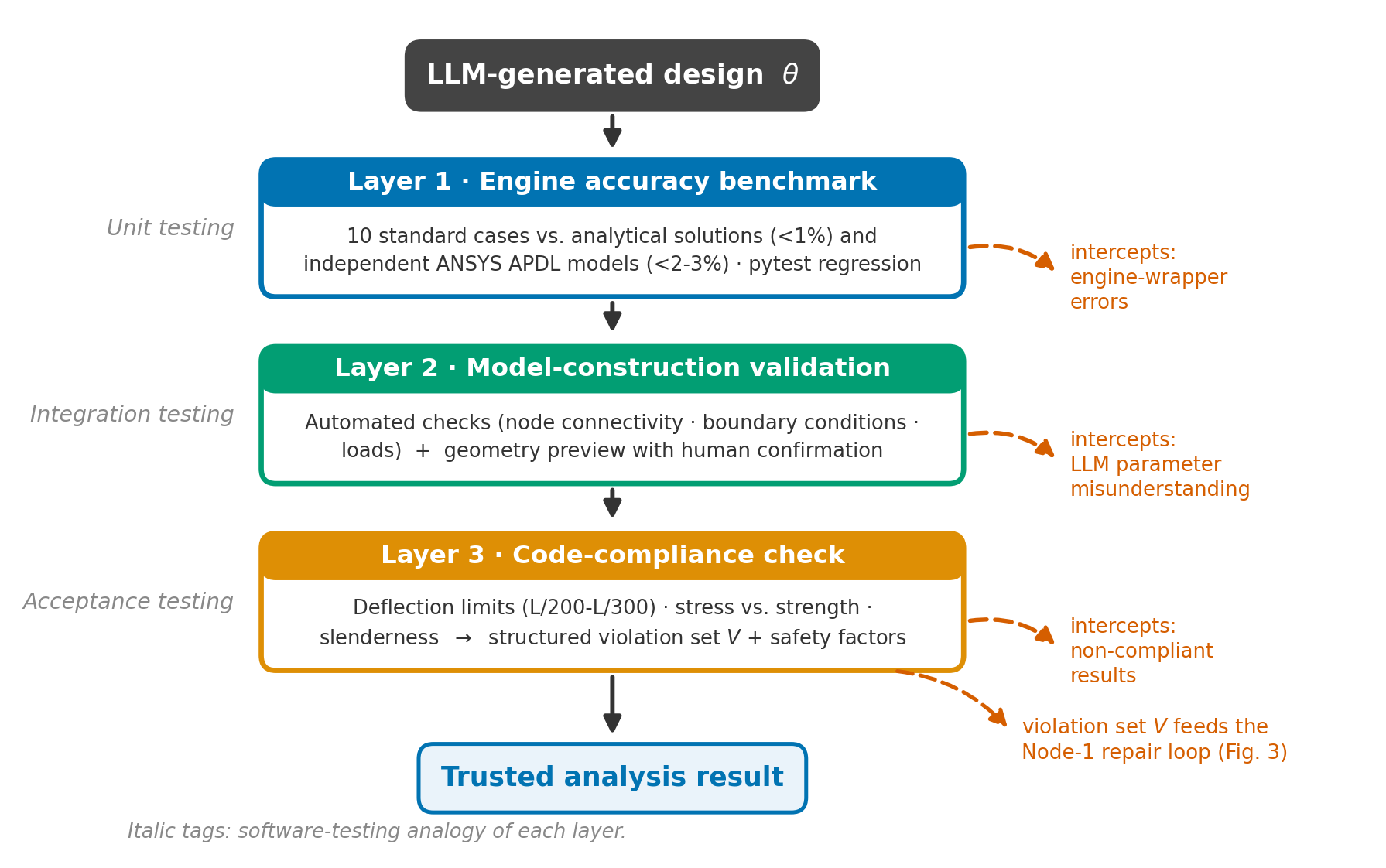}
\caption{Three-layer verification system (defence-in-depth against LLM hallucination).}
\label{fig:verify}
\end{figure}

In an LLM-driven design workflow, errors can enter at three levels\nrev{. They arise} in the low-level finite-element engine wrapper (which would invalidate every higher-level conclusion), in model construction when the LLM misreads parameters, and in code compliance at the result level. To match these, we build a three-layer verification system (Fig.~\ref{fig:verify}) that mirrors the unit-, integration- and acceptance-test levels of software engineering and can serve as a general verification template for LLM-driven engineering-computation systems.

\textbf{Layer 1: engine-accuracy benchmark.} Two standard cases are preset for each of the five structure types, giving ten regression tests. Where a theoretical analytical solution exists (e.g. simply supported and cantilever beams), the reference value is the mechanics-of-materials solution, and the OpenSeesPy relative error is required to be below 1\%; where the analytical solution is complex or unavailable (e.g. continuous beams, trusses, single-/double-storey frames), the reference value comes from an independent ANSYS APDL model \nrev{(its independence is what makes the comparison meaningful)}, with the tolerance relaxed to 2--3\% to absorb mesh-discretization error. All cases are bundled into a pytest regression suite, and any change at the engine layer must pass every assertion before it can enter the workflow. Results are reported in Section~\ref{sec:engine}.

\textbf{Layer 2: model-construction correctness.} Before each analysis, \texttt{\_validate\_model()} runs three checks\nrev{. These are} node connectivity (a non-zero node count), boundary-condition completeness (at least one fixed degree of freedom) and load-application correctness (with case-specific checks implemented per structure-type subclass). On top of this, FEAnalysisAgent draws a geometric preview before solving and shows the user the structure type, dimensions, supports and loads through human--machine interaction, starting computation only after confirmation. This layer targets a failure mode peculiar to LLMs\nrev{, namely parameter misreadings such as confusing span with total length or mistaking load direction,} and catches it before the costly finite-element solve, improving overall throughput.

\textbf{Layer 3: result code-compliance.} After analysis, \texttt{check\_code()} renders a static verdict against code limits\nrev{. These cover} maximum-deflection limits ($L/250$ for simply supported beams and trusses, $L/200$ for cantilever beams, $L/300$ for continuous beams and frames), maximum stress within the material design strength, slenderness limits for truss compression members, and so on. The output is a structured list of violation items $V$\nrev{, each carrying the violation type, actual value, limit and exceedance magnitude,} together with a safety factor for each checking dimension. $V$ carries the feedback signal to the first closed-loop node of Section~\ref{sec:dualnode}.

\subsection{Dual-node closed-loop feedback mechanism}\label{sec:dualnode}
\begin{figure}[htbp]
\centering
\includegraphics[width=0.95\linewidth]{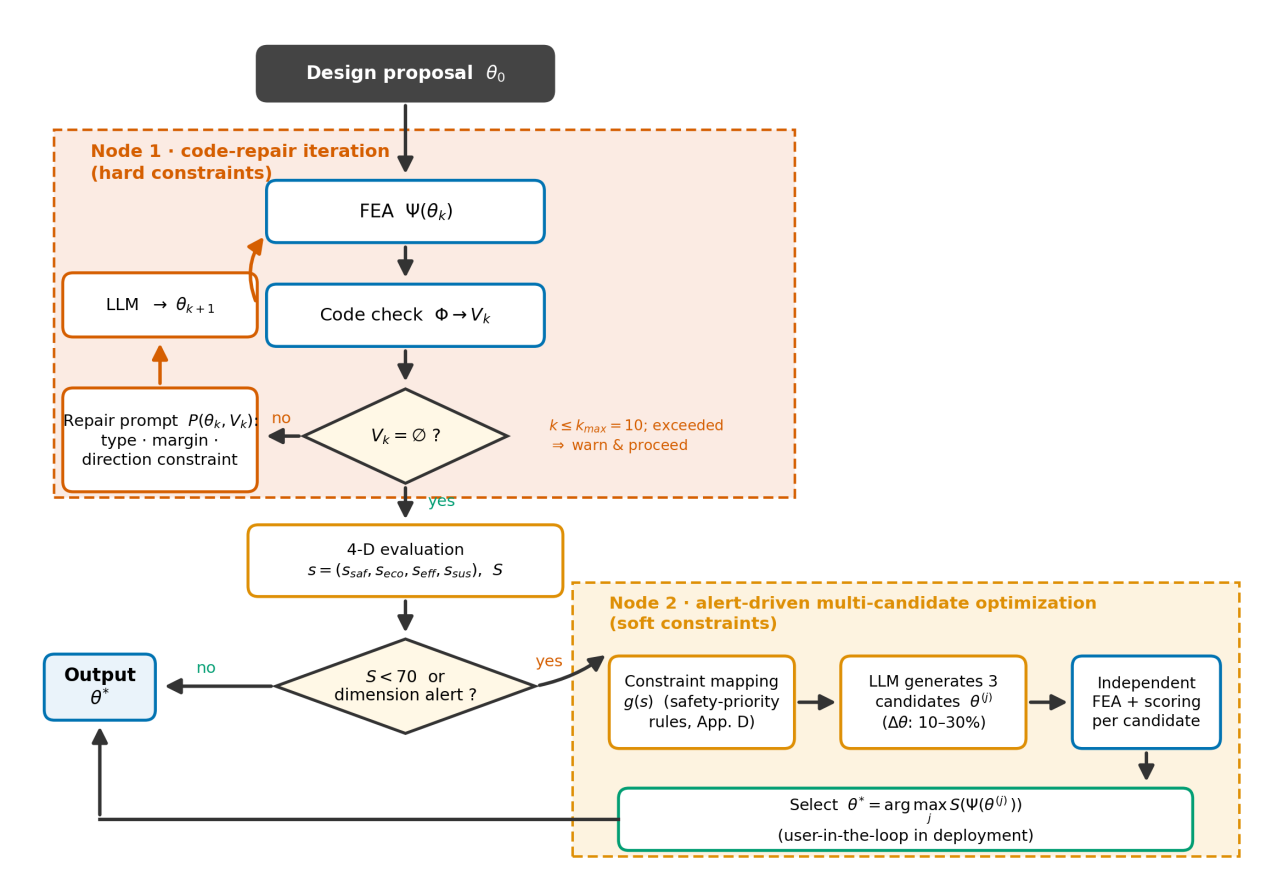}
\caption{Dual-node closed-loop feedback mechanism.}
\label{fig:loop}
\end{figure}

Let the design-parameter vector be $\theta$, the initial scheme $x_0$, the finite-element operator $\Psi$ and the code-checking operator $\Phi$. The violation-item set is then given by Eq.~\eqref{eq:violation}, and the closed loop chains two nodes in series, as shown in Fig.~\ref{fig:loop}.
\begin{equation}\label{eq:violation}
V = \Phi(\Psi(\theta))
\end{equation}

\textbf{Node~1: code-repair iteration (hard-constraint repair loop).} If $V\neq\varnothing$, a repair prompt $P_r(\theta,V)$ is constructed, listing item by item the violation type, actual value, limit and exceedance magnitude, together with a directional constraint (e.g. ``deflection exceeds the limit by 21\%; increase the section height or raise the material grade, and do not reduce the section''), which drives the LLM to iterate according to Eq.~\eqref{eq:iterate}:
\begin{equation}\label{eq:iterate}
\theta_{k+1} = \mathrm{LLM}\big(P_r(\theta_k, V_k)\big),\qquad k = 0,1,2,\ldots
\end{equation}
At each round, $\Psi$ and $\Phi$ are re-run until $V=\varnothing$ or $k$ reaches the cap $k_{\max}=10$. Writing the violations into the prompt in structured form, rather than letting the LLM retry freely, matters\nrev{. The} direction and magnitude of each exceedance act as an equivalent gradient, turning blind retries into directional repair. This parallels Self-Debug \cite{chen2024}, which injects interpreter errors into the prompt, except that here the feedback source is a physics solver rather than a code interpreter. In our experiments most cases converge within a few rounds (Section~\ref{sec:decomp}), so $k_{\max}=10$ leaves ample margin; if compliance is still not reached at $k_{\max}$, the system proceeds with an explicit warning and leaves the decision to the user.

\textbf{Node~2: warning-driven multi-scheme optimization (soft-constraint refinement loop).} Once code checking passes, the system runs the four-dimensional assessment of Section~\ref{sec:assess} to obtain a score vector $\mathbf{s}=(s_1,s_2,s_3,s_4)$ and a composite score $S$. The node fires when $S$ falls below the 70-point threshold or either the safety or the economy dimension falls below its warning threshold\nrev{. A} constraint-mapping function $g(\mathbf{s})$ turns each dimension's score band into parameter-adjustment rules (released with the code) and drives the LLM to generate three candidate schemes in sequence within the constraint space; each candidate independently undergoes finite-element analysis and four-dimensional assessment, and the best is chosen by Eq.~\eqref{eq:select} (in automated testing, the highest composite score).
\begin{equation}\label{eq:select}
x^{*} = \arg\max_{i} S\big(\Psi(x^{(i)})\big)
\end{equation}
The mapping is designed to keep the LLM from making unsound trade-offs across objectives. \nrev{Its main rules are as follows. When} $s_{\text{safety}}<75$, the section may not be reduced nor the material downgraded; when $s_{\text{economy}}<70$, the material may not be upgraded and the section may not shrink by more than 25\%; and when safety and economy are both low, safety takes priority and economy yields, so the LLM cannot buy economy at the expense of safety margin. Each candidate adjusts only the single weakest dimension, by 10--30\%, with span and load fixed, keeping the parameter search within a physically reasonable range. The need for this mechanism is tested by ablation A4 (Section~\ref{sec:ablation}).

The two nodes play distinct, sequential roles. Node~1 brings a non-compliant design into compliance and therefore matters mainly when the initial sections are clearly deficient; Node~2 then lifts the quality of an already-compliant scheme, giving a stable but bounded score gain. The contribution decomposition in Section~\ref{sec:decomp} tests this directly\nrev{. Score} history logs the score at three points \nrev{(initial, after Node~1 and after Node~2)} so the gain of each stage can be quantified separately.

\subsection{Four-dimensional quantitative assessment and RAG code tracing}\label{sec:assess}

\subsubsection{Four-dimensional quantitative assessment model}
The model spans four dimensions\nrev{, namely safety, economy, structural efficiency and sustainability}\nrev{. Safety} combines strength, stiffness and detailing checks; economy uses an optimal-utilization-band curve plus a material-usage index; structural efficiency combines stress-utilization level with utilization uniformity; and sustainability accounts for carbon-emission intensity per unit load-carrying capacity and material recyclability. The four are weighted into a 0--100 composite score mapped to grades A+ through D. Any code violation caps the composite score at 60, keeping the score consistent with the compliance verdict.

Dimension weights are set per structure type (e.g. a 45\% safety weight for cantilever beams and frames, a 30\% economy weight for trusses) through an auditable two-factor scoring--normalization procedure\nrev{. Each} type's failure-consequence severity $C$ and economic-optimization potential $E$ are scored on a 1--5 scale; the raw score follows Eq.~\eqref{eq:weight} ($\alpha=2$, $\beta=1.5$, with efficiency and sustainability fixed at baseline scores $R_3=4$, $R_4=3$), which is then normalized into the model weights and lightly tuned with engineering experience.
\begin{equation}\label{eq:weight}
R = \alpha C + \beta E + R_3 + R_4
\end{equation}
Because the procedure is transparent, reviewers and users can audit the weights; a quantitative analysis of how robust the main conclusions are to weight perturbation is left to future work.

\nrev{One caveat applies.} The composite score is at once an assessment metric and the optimization objective of Node~2, which raises the risk of a self-evaluation loop. For this reason, every quality conclusion in Section~\ref{sec:results} is reported alongside objective metrics that are independent of the scoring system \nrev{(code-compliance rate, number of violation items, minimum safety factor, material volume, and stress/deflection utilization)} as a cross-check.

\subsubsection{RAG code knowledge base}
Building on retrieval-augmented generation \cite{lewis2020}, the vector store is ChromaDB with text-embedding-ada-002 embeddings, currently holding GB~50010 (Code for Design of Concrete Structures) and GB~50017 (Standard for Design of Steel Structures). Documents are chunked at two levels\nrev{, first by heading and then, for chunks longer than 400 characters, at line breaks with a 50-character overlap between neighbours,} and tagged with Boolean labels by applicable structure type for filtered retrieval (falling back automatically to whole-base retrieval when the filter is empty). The RAG capability is injected into each evaluator as a Mixin\nrev{. A} detailing-check failure triggers retrieval, and the code number, clause number and a text excerpt are concatenated into a clause citation that is written into the violation record and carried through to the report, giving every diagnosis clause-level traceability. The effect of these citations on the repair result is tested by ablation A3 (Section~\ref{sec:ablation}).

\subsection{Input-robustness design and system implementation}\label{sec:impl}
Natural-language input brings boundary cases\nrev{ such as ambiguity, contradiction and incompleteness.} The system prompt of StructuralDesignAgent lays down explicit interaction rules for six such scenarios (ambiguous wording, conflicting parameters, multiple missing parameters, authorized completion, irrelevant input and cancellation intent)\nrev{. Ambiguous} wording must be clarified rather than guessed, conflicts must be listed in full for user confirmation, and two or more missing parameters are merged into a single templated query. This moves the boundary-handling logic from the code layer up to the prompt layer. Robustness results are reported in Section~\ref{sec:robust}.

\begin{figure}[htbp]
\centering
\includegraphics[width=0.95\linewidth]{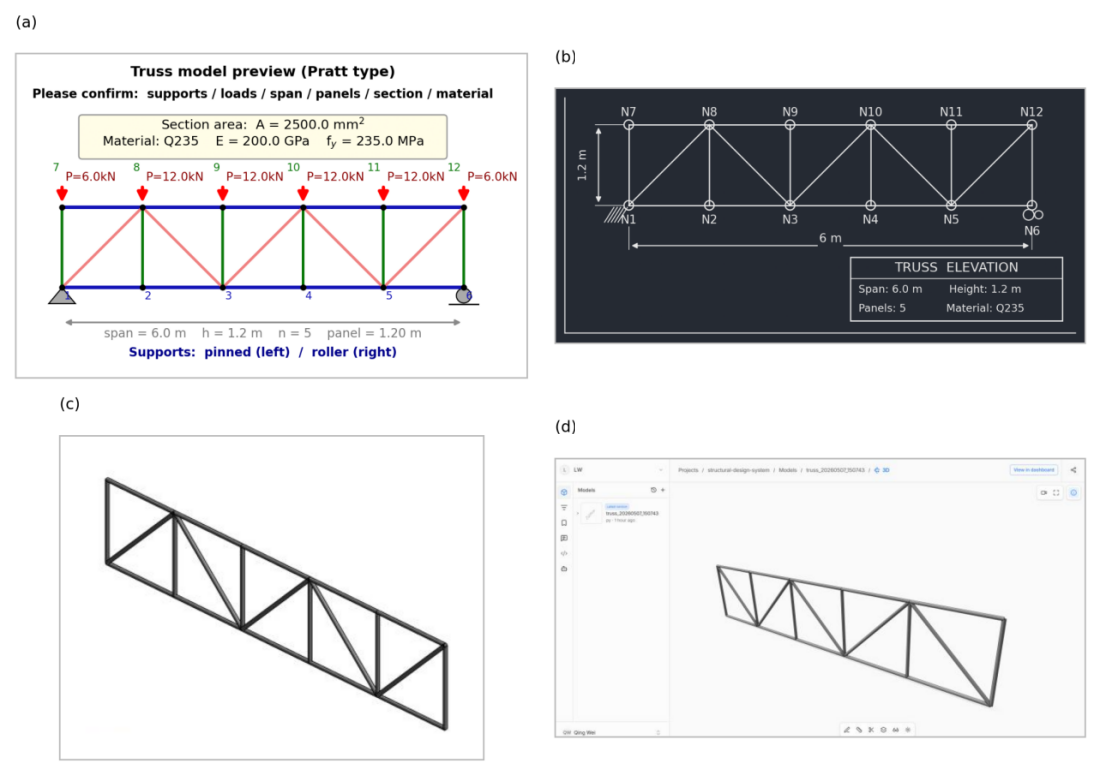}
\caption{System outputs of one truss design across four delivery channels: (a) pre-analysis geometry preview (Layer-2 human confirmation); (b) auto-generated CAD elevation drawing (DXF via ezdxf); (c) IFC4 export opened in Autodesk Revit (viewport; ifcopenshell); (d) browser-based 3-D collaboration (Speckle viewer).}
\label{fig:outputs}
\end{figure}

\begin{figure*}[htbp]
\centering
\includegraphics[width=\linewidth]{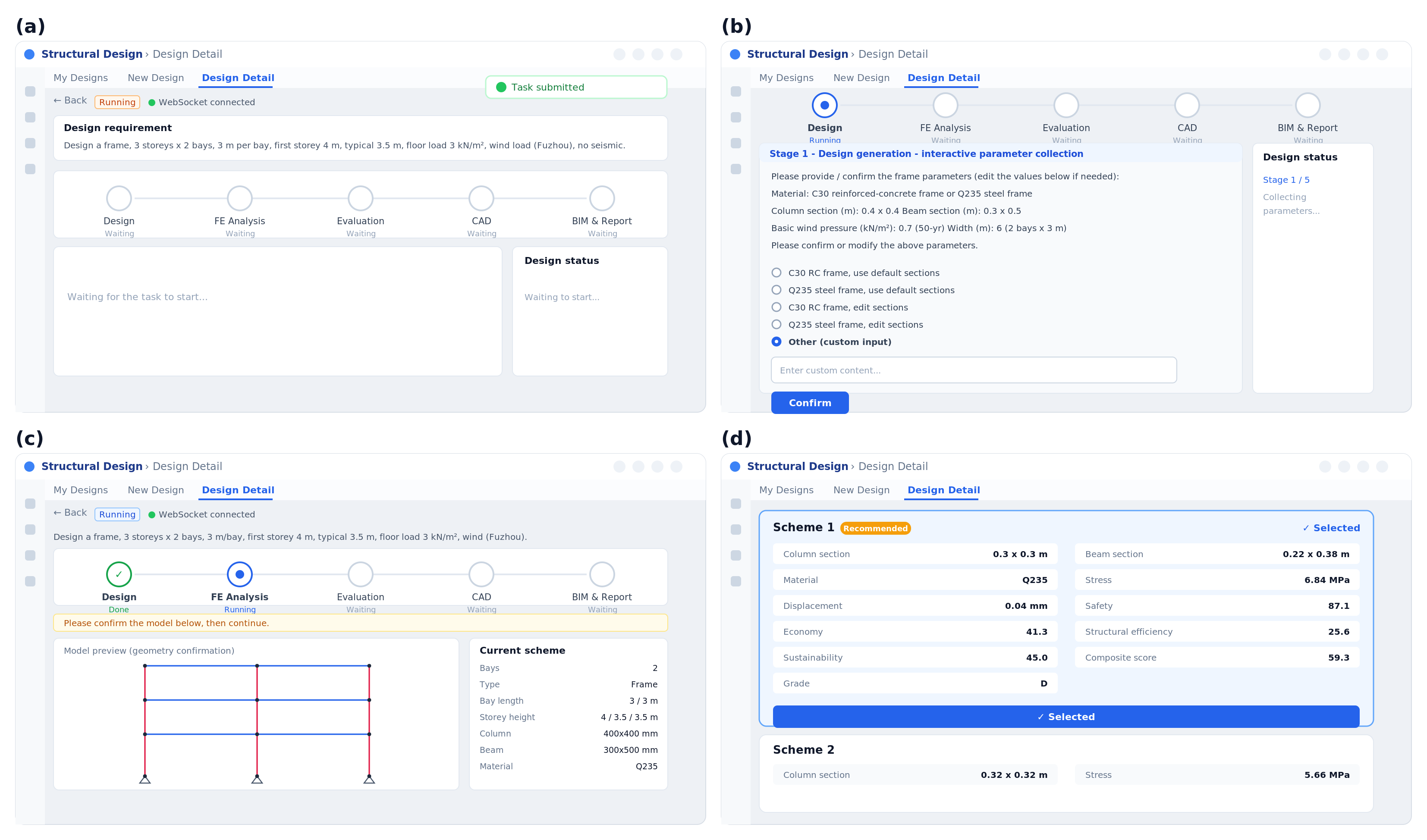}
\caption{\nrev{The browser-based design platform across the workflow: (a) a submitted natural-language design requirement, with the five-stage pipeline and a live WebSocket connection; (b) interactive collection and confirmation of design parameters; (c) the finite-element stage with geometric-preview confirmation and a live parameter panel; (d) four-dimensional evaluation of the candidate schemes for user selection.}}
\label{fig:webui}
\end{figure*}

The system runs as a web platform\nrev{. FastAPI} serves as the API gateway, and Celery with Redis forms an asynchronous task queue that decouples the long-running design workflow from the HTTP request; WebSocket with Redis Pub/Sub pushes stage progress in real time, and a single WebAskHuman tool encapsulates all human--machine waiting logic. The finite-element engine wraps OpenSeesPy 3.7.1 \cite{mckenna2011}, CAD drawings are produced as DXF via ezdxf, and BIM export supports IFC4 (ifcopenshell) and Speckle online 3D collaboration. Fig.~\ref{fig:outputs} shows one truss design delivered across four output channels, and \nrev{Fig.~\ref{fig:webui} shows the browser-based interface itself across the workflow, from natural-language requirement entry through real-time multi-stage progress, finite-element confirmation and four-dimensional evaluation.} Further implementation detail is omitted here; the complete code is released with the paper.

\section{Experimental design}
The experiments address four research questions. (RQ1) How much does the closed loop improve compliance and design quality over open-loop one-shot generation and weaker baselines? (RQ2) What does each of the two closed-loop nodes contribute, and where does each reach its limit? (RQ3) Is every verification component \nrev{(repair iteration, warning optimization, RAG citation, constraint rules)} necessary? (RQ4) Is the framework's performance decoupled from the underlying LLM?

\subsection{Environment and configuration}
\nrev{The hardware is} AMD Ryzen 7 6800H, 16~GB DDR4, Windows 11. \nrev{The software stack is} Python 3.12, OpenManus, OpenSeesPy 3.7.1, ezdxf 1.1.3 and ChromaDB. The backbone LLM is DeepSeek V4, sampled at temperature $=0$. At this setting the composite-score standard deviation of one-shot generation is exactly 0.0 across 5/10 repetitions, so generation is essentially deterministic; the repeated runs here therefore probe not sampling randomness but the process randomness of the closed-loop path once user decisions are defaulted\nrev{, the same source as the small per-case score fluctuation (SD of about 0--5 points) seen in the B4 closed loop.} This convention is held constant throughout the statistics.

\subsection{Test set}
The test set is a parametric family of 44 cases across five structure types. Within each type, cases are split by initial condition into initially under-sized sections (correction scenario, 23 cases) and initially compliant (refinement scenario, 21 cases), matching the role-division hypothesis of Section~\ref{sec:dualnode}. Table~\ref{tab:testset} lists the configurations.

\begin{table}[htbp]
\centering
\caption{Composition of the parametric test set ($N=44$).}
\label{tab:testset}
\small
\begin{tabularx}{\linewidth}{@{}L{0.20\linewidth} X C{0.10\linewidth} C{0.12\linewidth} C{0.12\linewidth}@{}}
\toprule
\textbf{Structure type} & \textbf{Parametric range} & \textbf{Variants} & \textbf{Repair} & \textbf{Refine} \\
\midrule
Simply supported beam & Span \{6,9,12,15\}\,m $\times$ load \{15,25,40\}\,kN/m & 12 & 8 & 4 \\
Cantilever beam & Length \{3,4,5,6\}\,m $\times$ material \{C30, Q345\} & 8 & 0 & 8 \\
Continuous beam & Spans \{2,3,4\} $\times$ span length \{5,6,8\}\,m & 9 & 0 & 9 \\
Truss & Span \{9,12,18\}\,m $\times$ panels \{6,8\} & 6 & 6 & 0 \\
Frame & Storeys \{2,3,4\} $\times$ bays \{1,2,3\} (partial) & 9 & 9 & 0 \\
\midrule
Total & --- & 44 & 23 & 21 \\
\bottomrule
\end{tabularx}
\end{table}

In the main experiment (E1), B4 (closed loop) and B3 (open loop) are each run 10 times per case (440 runs per configuration); in the baselines (E2), B1 and B2 are each run 5 times per case (220 runs per configuration); ablation (E3) runs each configuration 5 times on a representative subset of 9 cases; and the cross-model experiment (E4) runs on a representative subset of 5 cases. Human--machine-interaction nodes are scripted to default paths for reproducibility. At runtime, material volume, stress/deflection utilization and minimum safety factor are persisted to \texttt{result\_json}, which the objective-metric cross-checks of Section~\ref{sec:results} draw on. Every run completes with \texttt{status\,=\,success} (100\%).

\subsection{Baseline and ablation configurations}
\nrev{We compare four configurations. B1 (bare LLM, single prompt) asks the LLM in one prompt for the complete design parameters, with no tools and no verification, judged afterwards by our own system to give a compliance floor with no safeguard. B2 (single-agent ReAct) gives one agent all the tools but no orchestration or closed loop, which separates multi-agent division of labour from the closed-loop mechanism. B3 (open-loop full workflow) runs the full five-agent workflow with both closed-loop nodes switched off. B4 (closed-loop full workflow) is the framework proposed here.}

Each ablation removes a single component from B4: \textbf{A1} removes Node~1 (going straight to assessment on a violation) to expose compliance degradation; \textbf{A2} removes Node~2 to expose the lost quality gain; \textbf{A3} removes RAG, so violations carry no clause citation; and \textbf{A4} removes the constraint rules, letting Node~2 optimize freely, to see whether the LLM trades safety margin for economy unchecked.

\textbf{Cross-model generalization:} under the B4 configuration, the LLM is replaced with Claude Sonnet~4.6 and compared against DeepSeek V4 on a representative subset of the E1 test set.

\subsection{Metrics and statistical methods}
\textbf{Primary objective metrics:} code-compliance rate (proportion of runs whose violation set is empty at completion), mean number of violation items, minimum safety factor, material volume, and stress/deflection utilization and their coefficients of variation. \textbf{System self-assessment auxiliary metrics:} the four dimension scores and the composite score (cross-presented with the objective metrics; see Section~\ref{sec:assess}). \textbf{Process metrics:} Node-2 trigger rate, the three-node score trajectory of score history, and the two-node contribution decomposition. \textbf{Efficiency and cost metrics:} stage-wise time, token consumption and monetary cost.

We use a case-level paired design, comparing the same case across configurations. The Wilcoxon signed-rank test reports two-sided $p$-values, an effect size (matched-pairs rank-biserial correlation $r$) and 95\% confidence intervals; compliance-rate differences use the McNemar test; the significance level is 0.05, and multiple comparisons use Holm correction. The statistical-analysis pipeline (\texttt{stats\_pipeline.py}) is released with the paper.

\section{Results and discussion}\label{sec:results}

\subsection{Engine-accuracy benchmark}\label{sec:engine}
All ten standard cases pass the regression assertions (Table~\ref{tab:bench}). For the uniform- and concentrated-load cases of the simply supported and cantilever beams, the relative errors in maximum deflection, bending moment and stress against the mechanics-of-materials solution are all below 1\%; for the continuous beam, truss and single-/double-storey frames, the relative errors against independent ANSYS APDL models stay below 2--3\%. This establishes the trustworthiness of the computational engine for every subsequent experiment.

\begin{table}[htbp]
\centering
\caption{Engine-accuracy benchmark regression for the ten standard cases.}
\label{tab:bench}
\small
\begin{tabularx}{\linewidth}{@{}c L{0.22\linewidth} L{0.20\linewidth} X c c@{}}
\toprule
\textbf{No.} & \textbf{Structure type} & \textbf{Load case} & \textbf{Reference source} & \textbf{Tol.} & \textbf{Result} \\
\midrule
1 & Simply supported beam & Uniform load & Analytical solution & $<1\%$ & Pass \\
2 & Simply supported beam & Concentrated load & Analytical solution & $<1\%$ & Pass \\
3 & Cantilever beam & Uniform load & Analytical solution & $<1\%$ & Pass \\
4 & Cantilever beam & Concentrated load & Analytical solution & $<1\%$ & Pass \\
5 & Two-span continuous beam & Uniform load & Independent ANSYS APDL & $<2\%$ & Pass \\
6 & Two-span continuous beam & Concentrated load & Independent ANSYS APDL & $<2\%$ & Pass \\
7 & Truss & Nodal uniform load & ANSYS APDL / analytical & $<2\%$ & Pass \\
8 & Truss & Mid-span concentrated & ANSYS APDL / analytical & $<2\%$ & Pass \\
9 & Single-storey frame & Nodal load & Independent ANSYS APDL & $<3\%$ & Pass \\
10 & Double-storey frame & Nodal load & Independent ANSYS APDL & $<3\%$ & Pass \\
\bottomrule
\end{tabularx}
\end{table}

To confirm the reference values themselves, we re-implemented cases 1--8 from scratch in OpenSeesPy using third-party code; the results appear in Table~\ref{tab:repro}. For the beam cases the maximum error stays below 0.2\%. For the two truss cases we report a stress-reproduction check only: the maximum member stress matches the reference value exactly under both hand calculation and numerical confirmation, whereas deflection depends on the specific web-member arrangement (Pratt type) and is not compared here, which is why the deflection entries are left blank. The reproduction script (\texttt{benchmark\_repro.py}) is released alongside the paper.

\begin{table}[htbp]
\centering
\caption{Independent reproduction of the standard cases.}
\label{tab:repro}
\scriptsize
\setlength{\tabcolsep}{4pt}
\begin{tabularx}{\linewidth}{@{}c X r r r r r c c@{}}
\toprule
\textbf{No.} & \textbf{Case} & $\boldsymbol{\delta_{\text{calc}}}$ & $\boldsymbol{\delta_{\text{ref}}}$ & $\boldsymbol{M_{\text{calc}}}$ & $\boldsymbol{\sigma_{\text{calc}}}$ & $\boldsymbol{\sigma_{\text{ref}}}$ & \textbf{Max err.} & \textbf{Verdict} \\
 &  & (mm) & (mm) & (kN\,m) & (MPa) & (MPa) &  &  \\
\midrule
1 & Simply supported -- uniform & 1.0417 & 1.0417 & 45.00 & 2.500 & 2.500 & 0.00\% & Pass \\
2 & Simply supported -- concentrated & 0.8333 & 0.8333 & 45.00 & 2.500 & 2.500 & 0.00\% & Pass \\
3 & Cantilever -- uniform & 0.6250 & 0.6250 & 45.00 & 2.500 & 2.500 & 0.00\% & Pass \\
4 & Cantilever -- concentrated & 1.1111 & 1.1111 & 60.00 & 3.333 & 3.333 & 0.00\% & Pass \\
5 & Two-span continuous -- uniform & 0.2083 & 0.2085 & 31.25 & 1.736 & 1.736 & 0.10\% & Pass \\
6 & Two-span continuous -- concentrated & 0.2157 & 0.2160 & 28.12 & 1.562 & 1.563 & 0.16\% & Pass \\
7 & Truss -- nodal uniform & --- & 0.2470 & --- & 4.000 & 4.000 & 0.00\%$^{\sigma}$ & Stress pass \\
8 & Truss -- mid-span concentrated & --- & 0.1330 & --- & 2.000 & 2.000 & 0.00\%$^{\sigma}$ & Stress pass \\
\bottomrule
\end{tabularx}
\end{table}

\subsection{Closed-loop vs. open-loop comparison}
Table~\ref{tab:overall} summarizes the four configurations over the 44 cases. The three open-loop configurations (B1, B2, B3) behave almost identically\nrev{, all at} a 56.8\% compliance rate and a composite score of 63.7--63.8, with no significant pairwise difference ($p>0.4$, Table~\ref{tab:sig}). This null result is itself informative\nrev{. Simply} adding agents or layering on ReAct does not lift performance; the gain comes from the verification-driven closed loop itself. With the closed loop in place (B4), the compliance rate climbs to 98.6\%, the composite score to 71.4, material usage falls from 1.423 to 1.341~m\textsuperscript{3}, and the mean minimum safety factor falls from about 10.4 to 6.7. The pattern suggests that the open-loop baseline reaches compliance by over-sizing sections, whereas the closed loop trims the redundant margin while staying compliant, improving compliance and economy at once. B4's edge over all three baselines is highly significant, with effect size $r=0.85$ and a McNemar test of $p<10^{-5}$ on compliance rate.

\begin{table}[htbp]
\centering
\caption{Overall performance comparison of the four configurations.}
\label{tab:overall}
\small
\begin{tabularx}{\linewidth}{@{}c X c c X c c@{}}
\toprule
\textbf{Cfg.} & \textbf{Description} & \textbf{Runs} & \textbf{Compl.} & \textbf{Composite (mean$\pm$SD)} & \textbf{Mat. (m\textsuperscript{3})} & \textbf{Min. SF} \\
\midrule
B1 & Bare LLM, single prompt & 220 & 56.8\% & 63.7$\pm$8.6 & 1.423 & 10.36 \\
B2 & Single-agent ReAct & 220 & 56.8\% & 63.8$\pm$8.5 & 1.423 & 10.35 \\
B3 & Multi-agent open loop & 440 & 56.8\% & 63.8$\pm$8.5 & 1.423 & 10.35 \\
B4 & Verification-driven closed loop (this work) & 440 & 98.6\% & 71.4$\pm$5.9 & 1.341 & 6.72 \\
\bottomrule
\end{tabularx}
\end{table}

\begin{table}[htbp]
\centering
\caption{Paired significance tests.}
\label{tab:sig}
\small
\begin{tabularx}{\linewidth}{@{}l c c c c c c X@{}}
\toprule
\textbf{Comparison} & \textbf{mean(A)} & \textbf{mean(B)} & \textbf{$W$} & \textbf{$p$(Holm)} & \textbf{$r$} & \textbf{McNemar $p$} & \textbf{Conclusion} \\
\midrule
B4 vs B1 & 71.4 & 63.7 & 9 & $6.3\!\times\!10^{-7}$ & 0.85 & $7.6\!\times\!10^{-6}$ & B4 superior (very large) \\
B4 vs B2 & 71.4 & 63.8 & 9 & $5.3\!\times\!10^{-7}$ & 0.85 & $7.6\!\times\!10^{-6}$ & B4 superior \\
B4 vs B3 & 71.4 & 63.8 & 9 & $4.2\!\times\!10^{-7}$ & 0.85 & $7.6\!\times\!10^{-6}$ & B4 superior \\
B1 vs B2 & 63.7 & 63.8 & 9 & 0.42 & 0.25 & 1.00 & No sig. difference \\
B1 vs B3 & 63.7 & 63.8 & 10 & 0.42 & 0.11 & 1.00 & No sig. difference \\
B2 vs B3 & 63.8 & 63.8 & 12 & 0.40 & 0.25 & 1.00 & No sig. difference \\
\bottomrule
\end{tabularx}
\end{table}

\begin{figure}[htbp]
\centering
\includegraphics[width=0.95\linewidth]{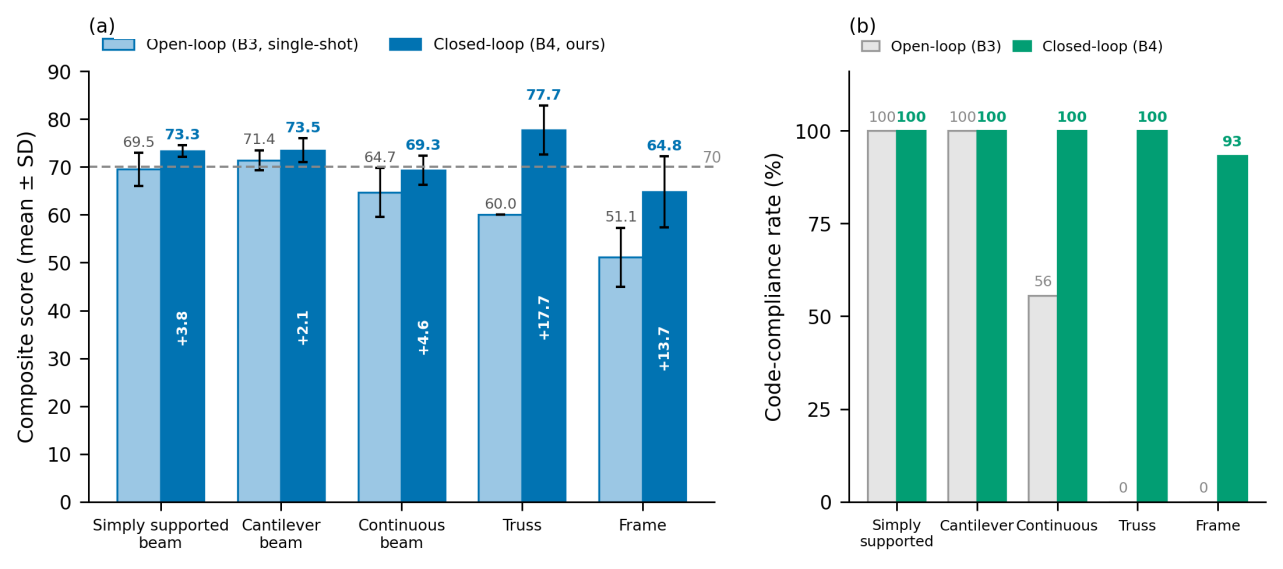}
\caption{Open-loop (B3) vs. closed-loop (B4) by structure type: (a) composite score (mean $\pm$ SD); (b) code-compliance rate.}
\label{fig:bytype}
\end{figure}

Broken down by structure type (Table~\ref{tab:bytype}, Fig.~\ref{fig:bytype}), the distribution of gains is clear. For types that start compliant \nrev{(simply supported and cantilever beams),} the open loop already passes at 100\%, and the closed loop only fine-tunes quality. The closed loop's value concentrates on continuous beams, trusses and frames, whose initial sections are typically insufficient\nrev{. The} truss composite score rises from 60.0 to 77.7 and the frame from 51.1 to 64.8, while their compliance rates rise from 0\% to 100\% and 93.3\%, respectively.

\begin{table}[htbp]
\centering
\caption{Open-loop (B3) vs. closed-loop (B4) by structure type.}
\label{tab:bytype}
\small
\begin{tabularx}{\linewidth}{@{}X c c c c c@{}}
\toprule
\textbf{Structure type} & \textbf{B3 compl.} & \textbf{B3 score} & \textbf{B4 compl.} & \textbf{B4 score} & \textbf{Gain} \\
\midrule
Simply supported beam & 100\% & 69.5 & 100\% & 73.3 & +3.8 \\
Cantilever beam & 100\% & 71.4 & 100\% & 73.5 & +2.1 \\
Continuous beam & 55.6\% & 64.7 & 100\% & 69.3 & +4.6 \\
Truss & 0\% & 60.0 & 100\% & 77.7 & +17.7 \\
Frame & 0\% & 51.1 & 93.3\% & 64.8 & +13.7 \\
\bottomrule
\end{tabularx}
\end{table}

\begin{figure}[htbp]
\centering
\includegraphics[width=0.78\linewidth]{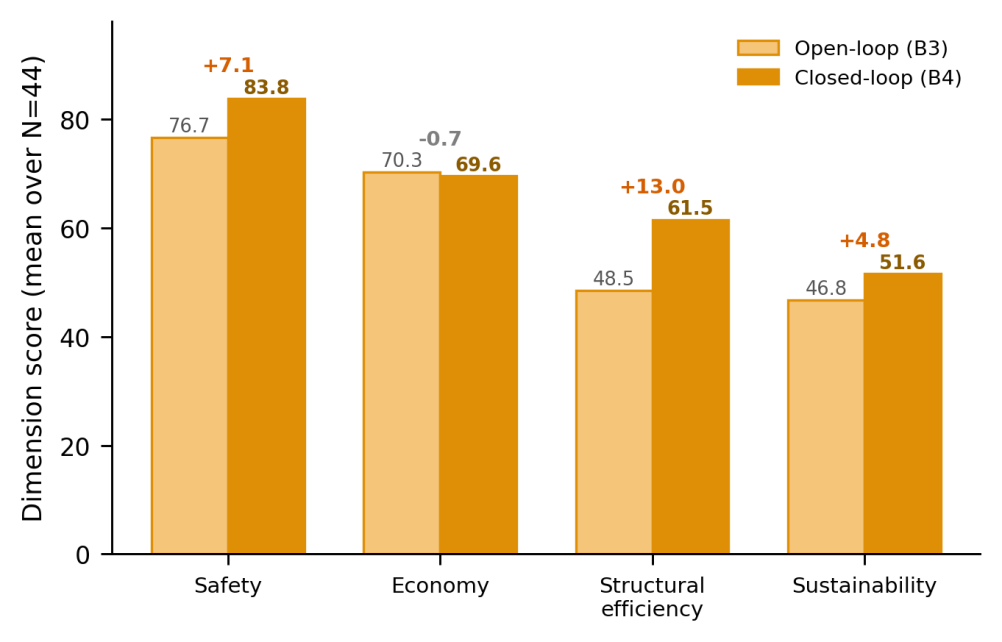}
\caption{Dimension-wise score comparison, open-loop vs. closed-loop.}
\label{fig:dim}
\end{figure}

By assessment dimension (Fig.~\ref{fig:dim}), the closed-loop gain lifts structural efficiency from 48.5 to 61.5 and safety from 76.7 to 83.8, with a slight rise in sustainability and economy essentially flat. Read together with the drop in material usage, economy is held steady while consuming less material.

\subsection{Closed-loop contribution decomposition and capability boundary}\label{sec:decomp}
\begin{figure}[htbp]
\centering
\includegraphics[width=0.8\linewidth]{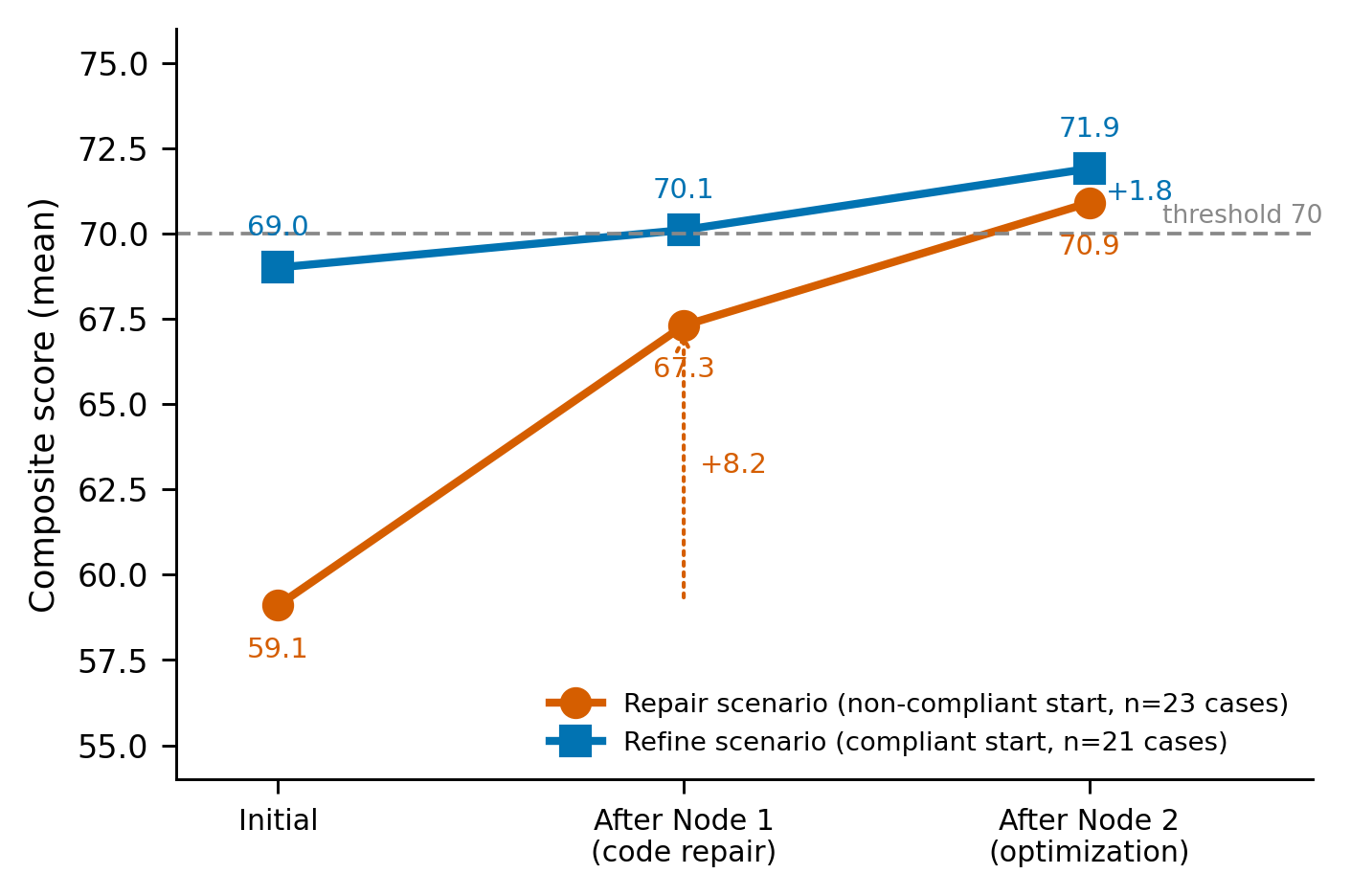}
\caption{Score trajectories across the two feedback nodes, repair vs. refine.}
\label{fig:traj}
\end{figure}

Using the score-history data, we decompose the closed-loop gain by scenario into the contributions of Node~1 (code repair) and Node~2 (warning optimization), as shown in Table~\ref{tab:decomp} and Fig.~\ref{fig:traj}. The division of labour in the two scenarios matches the hypothesis of Section~\ref{sec:dualnode} closely.

\begin{table}[htbp]
\centering
\caption{Dual-node contribution decomposition.}
\label{tab:decomp}
\small
\begin{tabularx}{\linewidth}{@{}X c c c c c@{}}
\toprule
\textbf{Scenario} & $\boldsymbol{s_0}$ & $\boldsymbol{s_1}$ & $\boldsymbol{s_2}$ & \textbf{Node 1} & \textbf{Node 2} \\
\midrule
Repair (initially non-compliant, 23 cases) & 59.1 & 67.3 & 70.9 & +8.2 & +3.6 \\
Refine (initially compliant, 21 cases) & 69.0 & 70.1 & 71.9 & +1.1 & +1.8 \\
\bottomrule
\end{tabularx}
\end{table}

In the repair scenario, of the +11.8-point total gain, Node~1 supplies +8.2 points \nrev{(it carries the feasibility repair)} and Node~2 adds a further +3.6 points of quality refinement. In the refinement scenario the initial scheme is already compliant, Node~1 is barely activated, and the +1.8-point gain comes almost entirely from Node~2. As a control, the two-node contributions are +0.0 for all three open-loop configurations (B1/B2/B3), confirming that these increments stem from the closed loop rather than from other factors.

Even after the full closed loop, the frame case still scores only 64.8 with a 93.3\% compliance rate (Table~\ref{tab:bytype}). The reason is that its structural-efficiency score is bounded by the initial topology\nrev{. The} 10--30\% section adjustments allowed by Node~2 leave the topology unchanged and cannot improve force uniformity. A parameter-level closed loop, then, can repair parameter-level defects and refine compliant schemes, but it cannot substitute for a topology-level redesign. Recognizing this boundary is essential to using the framework correctly and marks a clear direction for future work.

The residual non-compliance is worth examining directly, because in a safety-critical setting what the system does when it fails matters as much as its success rate. The 98.6\% figure means that about six of the 440 closed-loop runs finish non-compliant, and all of them are frame cases; the two beam types and the truss reach 100\%. These failures share the mechanism just described: the deficiency originates in the initial topology, so Node~1 iterates but cannot bring stiffness or force-flow uniformity within limits by section change alone, and the loop stops at the cap $k_{\max}=10$. Crucially, the system does not silently pass such a scheme. It terminates with an explicit non-compliance warning, keeps the residual violation-item list and the governing safety factor visible in the report, marks the scheme as failed rather than delivered, and leaves the decision to the engineer. In other words, the framework fails loudly rather than quietly, surfacing exactly the cases that call for manual judgement or a topology-level redesign. This behaviour is consistent with the trustworthy positioning of the work: the value of an external verifier lies not only in raising the pass rate but in refusing to certify what it cannot verify.

\subsection{Ablation and cross-model results}\label{sec:ablation}
\begin{figure}[htbp]
\centering
\includegraphics[width=0.95\linewidth]{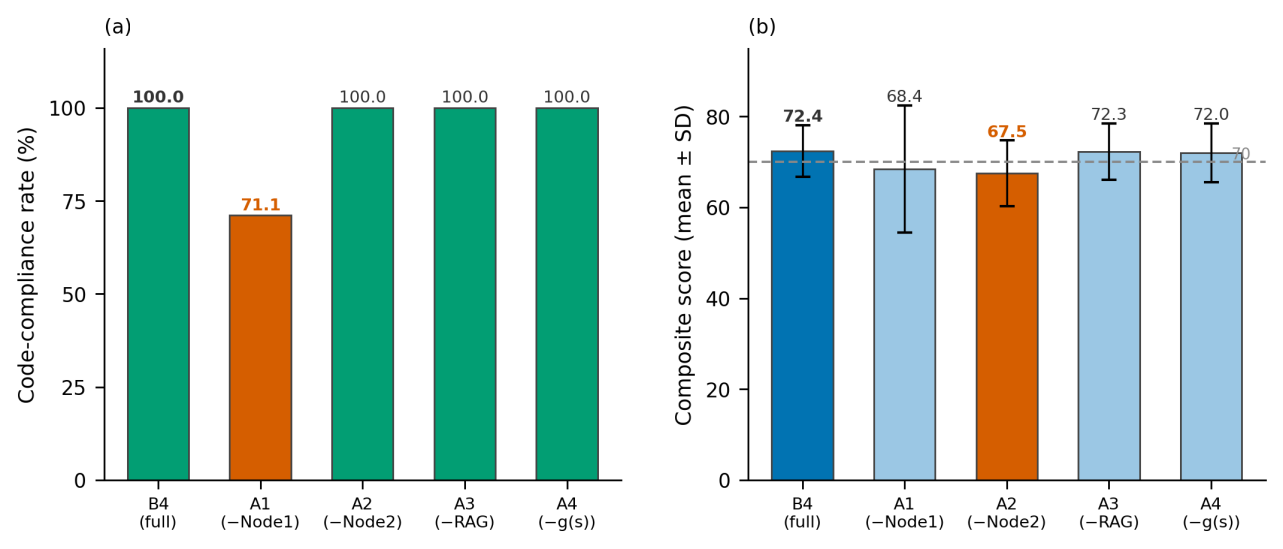}
\caption{Ablation results (B4 vs. A1--A4): (a) code-compliance rate; (b) composite score (mean $\pm$ SD).}
\label{fig:ablation}
\end{figure}

We remove components one at a time on a representative subset of 9 cases; results appear in Table~\ref{tab:ablation} and Fig.~\ref{fig:ablation}. On this subset, full B4 reaches 100\% compliance with a score of 72.4$\pm$5.7.

\begin{table}[htbp]
\centering
\caption{Ablation experiments.}
\label{tab:ablation}
\small
\begin{tabularx}{\linewidth}{@{}c X c c c c@{}}
\toprule
\textbf{Cfg.} & \textbf{Removed component} & \textbf{Compl.} & \textbf{Composite (mean$\pm$SD)} & \textbf{$p$(Holm)} & \textbf{$r$} \\
\midrule
B4 & --- (full) & 100.0\% & 72.4$\pm$5.7 & --- & --- \\
A1 & Remove Node 1 & 71.1\% & 68.4$\pm$14.0 & 1.00 & $-0.33$ \\
A2 & Remove Node 2 & 100.0\% & 67.5$\pm$7.2 & 0.039 & $-1.00$ \\
A3 & Remove RAG code retrieval & 100.0\% & 72.3$\pm$6.2 & 1.00 & $-0.33$ \\
A4 & Remove $g(\mathbf{s})$ optimization constraints & 100.0\% & 72.0$\pm$6.5 & 1.00 & $-0.33$ \\
\bottomrule
\end{tabularx}
\end{table}

The findings are as follows. \nrev{For \textit{A1}, compliance} collapses from 100\% to 71.1\% and the score standard deviation widens from 5.7 to 14.0, a sharp loss of output stability that marks Node~1 as the linchpin of compliance and consistency. \nrev{For \textit{A2}, the score} falls from 72.4 to 67.5 ($p_{\text{Holm}}=0.039$, $r=-1.00$), establishing that Node~2 makes a reliable contribution to design quality. \nrev{For \textit{A3} and \textit{A4}, neither} materially affects the final composite score on this subset; their value lies in code-citation traceability and a sensible optimization path rather than in the score, and they are best judged against the interpretability objective rather than a single number.

\begin{table}[htbp]
\centering
\caption{Cross-model generalization.}
\label{tab:crossmodel}
\small
\begin{tabularx}{\linewidth}{@{}X c c c c@{}}
\toprule
\textbf{Backbone model} & \textbf{Runs} & \textbf{Compl.} & \textbf{Composite (mean$\pm$SD)} & \textbf{vs DeepSeek $p$} \\
\midrule
DeepSeek V4 (baseline) & 50 & 100.0\% & 72.9$\pm$5.2 & --- \\
Claude Sonnet 4.6 & 15 & 100.0\% & 71.5$\pm$7.8 & 0.44 \\
\bottomrule
\end{tabularx}
\end{table}

\begin{figure}[htbp]
\centering
\includegraphics[width=0.62\linewidth]{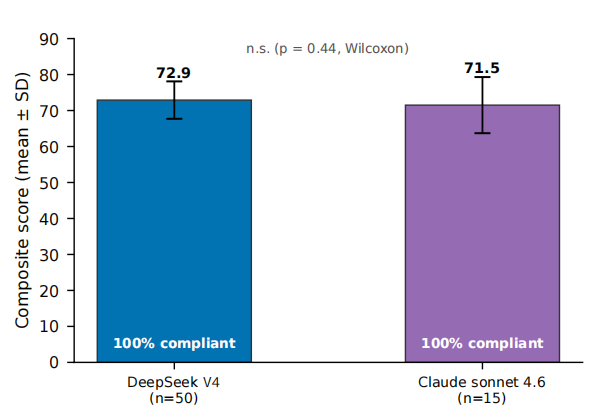}
\caption{Cross-model generalization of the closed loop: DeepSeek V4 vs. Claude Sonnet 4.6.}
\label{fig:crossmodel}
\end{figure}

Replacing the LLM from DeepSeek V4 with Claude Sonnet~4.6, the closed-loop framework still reaches 100\% compliance with a composite score of 71.5$\pm$7.8, a gap of only 1.4 points and a Wilcoxon $p=0.44$ (Table~\ref{tab:crossmodel}, Fig.~\ref{fig:crossmodel}). This test has limited statistical power (a five-case subset, $n=15$ against $n=50$), so we read the result as no detectable difference between the two backbones rather than as proven model-independence; it is consistent with the central premise of verification-driven design, that compliance is here underwritten by the external verifier rather than by model capability, but a broader multi-model study is needed to establish the claim firmly and is left to future work. Qwen-Max could not be included because its tool-call output format was not parsed by our design-scheme extractor, a protocol-compatibility issue rather than a limitation of the method.

\subsection{Efficiency and cost}
\begin{figure}[htbp]
\centering
\includegraphics[width=0.95\linewidth]{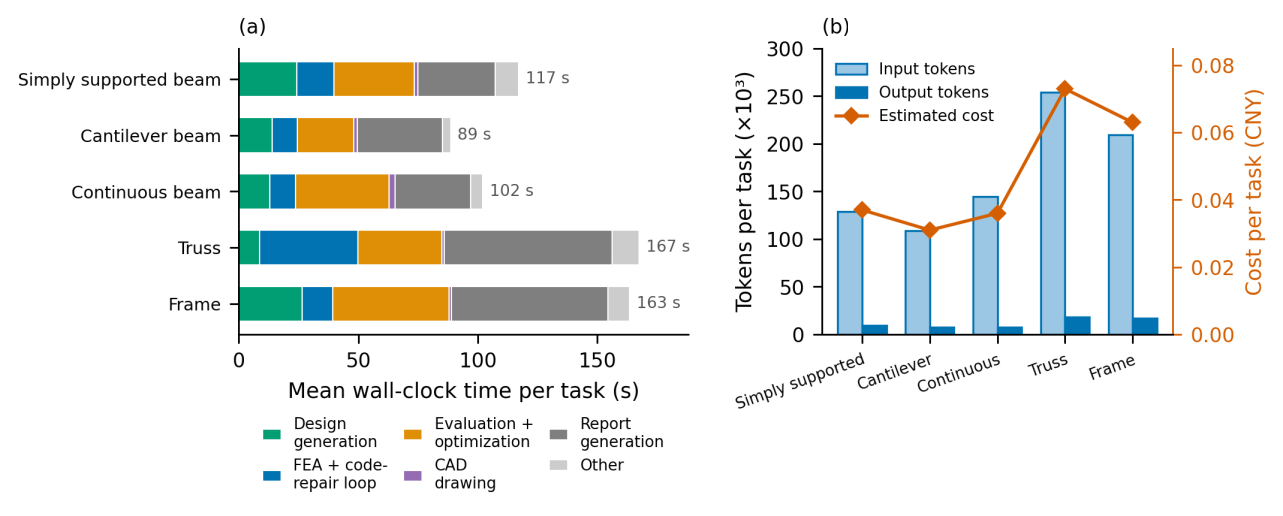}
\caption{Efficiency and cost: (a) stage-wise time breakdown; (b) token usage and cost per task.}
\label{fig:cost}
\end{figure}

Across 25 complete web-side interaction tests, total time per task ranges from 65 to 230~s, averaging 127.69~s, distributed as report generation 46.98~s, assessment and multi-scheme optimization 36.03~s, finite-element analysis 18.14~s, design-scheme generation 17.29~s, and CAD drawing under 3~s (Fig.~\ref{fig:cost}a). Including real human--machine interaction, an end-to-end design takes about 5--12 minutes\nrev{, roughly an order of magnitude faster than the two hours of manual effort reported in the literature} \cite{liang2025}.

Token consumption is strongly input-biased\nrev{, with 169{,}251 input tokens against 11{,}957 output on average, a ratio of about 14.2:1. This is} a consequence of the context that accumulates as structured data are passed explicitly between stages (Fig.~\ref{fig:cost}b). At list pricing the cost ceiling averages CNY~0.193 per task; with a measured 87.5\% cache-hit rate the actual cost falls to about CNY~0.048 per task, or CNY~1.20 over 25 runs, matching the CNY~1.19 actually billed. Cost varies mainly with the execution path\nrev{. The} truss case, which triggers both the repair loop and multi-scheme optimization, consumes the most tokens at 272{,}544. Expressed as a price-independent invariant, the closed loop exchanges a $2$--$3\times$ compute overhead for a gain of about $42$ percentage points in compliance (56.8\% to 98.6\%); this trade ratio is a property of the method, whereas its monetary value (here negligible, at about CNY~0.05 per task) tracks API pricing and may change over time.

\subsection{Input robustness}\label{sec:robust}
We designed 33 test cases spanning six boundary scenarios\nrev{, namely ambiguous wording, conflicting parameters, missing key parameters, irrelevant input, user cancellation and authorized completion.} Three initially failed, owing to clarification loops exceeding the round limit and a misread cancellation intent; after targeted fixes, all 33 passed, showing that prompt-layer boundary rules can reliably govern agent behaviour under abnormal input. Per-category outcomes are released with the code.

\subsection{Discussion}
Against the three gaps of Section~\ref{sec:gaps}, the closed loop lifts an open-loop baseline from 56.8\% to 98.6\% compliance and, through the score-history decomposition, shows where the gain arises (Node~1 supplies about 70\% in correction scenarios, Node~2 the quality refinement; Section~\ref{sec:decomp}); the three-layer system verifies engine, model and result rather than comparing once after the fact; and every violation is traced to a GB~50010 or GB~50017 clause under transparent, auditable weights. Relative to the closest multi-agent studies, which establish that such systems can design and verify after the fact \cite{chen2025,liang2025}, our emphasis is orthogonal: we quantify what a closed verification loop adds on the same agents (about 42 percentage points of compliance for a 2--3$\times$ compute overhead), and the null result across B1--B3 attributes this gain to the loop rather than to more agents or ReAct (Section~\ref{sec:ablation}).

In practice the framework is a verifiable assistant, not an autonomous designer. It is best applied where the topology is already reasonable and the open question is member sizing and code compliance, since the loop optimizes parameters under a fixed topology; and because it fails loudly rather than silently, its explicit non-compliance warnings should be treated as triage signals for manual review or topology redesign. A qualified engineer retains final judgement, auditing each decision through the clause-level citations, and remains responsible for the delivered design.

\section{Limitations}
Three limitations bound our conclusions. First, structural coverage is limited to five planar linear-elastic systems, and the loop adjusts parameters under a fixed topology, so defects originating in the topology cannot be repaired (Section~\ref{sec:decomp}). Second, the assessment carries methodological risk: the weights, though transparent, are partly subjective, and the composite score is both a metric and Node~2's objective; we mitigate this by cross-checking against independent quantities (compliance rate, safety factor, material usage) but cannot remove it, and a weight-sensitivity analysis and a blind comparison by registered engineers are left to future work. Third, intrinsic LLM limits persist: the knowledge base holds only two standards and RAG fires only at assessment, so it mainly adds traceability, and because the method combines prompting with external verification rather than fine-tuning, hallucination is suppressed but not eliminated, while long-context instruction-following still warrants systematic evaluation.

\section{Conclusions}
\nrev{Trustworthiness is the key to bringing LLM-driven structural design into practice. This paper proposes a closed-loop multi-agent framework whose feedback comes not from the model's self-reflection but from two external verifiers, namely finite-element analysis and code checking. The three-layer verification first guarantees the trustworthiness of the computational foundation (engine-benchmark error within 3\%). The dual-node closed loop then transforms violations and score gaps into repair constraints executable by the LLM. The RAG knowledge base grounds every violation diagnosis in a specific clause.}

\nrev{On five structure types and 44 parametric cases, the closed loop raises the code-compliance rate from 56.8\% (open loop) to 98.6\% and the composite score from 63.8 to 71.4 (relative improvement 11.9\%, $p<10^{-6}$, $r=0.85$), while reducing material usage by about 5.8\%. The score-history decomposition supplements the previously missing process perspective. Code-repair iteration addresses feasibility and contributes about 70\% of the improvement in repair scenarios, while warning optimization is responsible for refinement. Ablation confirms that neither node is dispensable. Removing Node~1 reverts the compliance rate to 71\% and removing Node~2 significantly lowers quality. The cross-model experiment further shows that the conclusions remain unchanged after replacing DeepSeek with Claude, with compliance guaranteed by the external verifier rather than a specific model. The frame cases also reveal the boundary of the method. When the defect originates from the structural topology, the parameter-level closed loop is powerless. At about 0.05~CNY and roughly two minutes per task, the cost of the closed loop is negligible in engineering terms.}

\nrev{Future work will proceed along three directions. The first extends the closed loop from parameter adjustment to topology modification, so that generative topology exploration also enters this verification loop. The second moves code retrieval forward into the design-generation and code-checking stages and introduces version management for the code library. The third explores the feasibility of migrating this three-layer verification system to other safety-critical domains.}

\printcredits

\section*{Declaration of competing interest}
The authors declare that they have no known competing financial interests or personal relationships that could have appeared to influence the work reported in this paper.

\section*{Acknowledgements}
This research was funded by the National Natural Science Foundation of China (Grant No. 52278295) and the Natural Science Foundation of Fujian Province (Grant No. 2024J01357).

\section*{Data and code availability}
\nrev{All source code supporting the conclusions of this paper is provided in a public GitHub repository under the MIT license (\url{https://github.com/fzuKilo/structural-design-system}). The release includes the} finite-element analysers for the five structure types, the RAG code-knowledge-base construction scripts, the 44 parametric test cases, the four configurations (B1--B4), the raw results of the ablation and cross-model experiments (\texttt{results\_*.jsonl}), the statistical-analysis pipeline (\texttt{stats\_pipeline.py}), the independent reproduction script (\texttt{benchmark\_repro.py}) and all plotting scripts. The repository contains the complete web-platform implementation, the multi-agent orchestration (PlanningFlow) and the five specialized agent implementations.

\section*{Declaration of generative AI and AI-assisted technologies in the manuscript preparation process}
During the preparation of this work, the authors used Anthropic Claude in order to improve manuscript clarity, conciseness, and grammar. After using this tool, the authors reviewed and edited the content as needed and take full responsibility for the content of the published paper.

\end{document}